\documentclass[aps,pre,superscriptaddress,twocolumn,amsmath,amssymb,showpacs]{revtex4}
\usepackage{graphicx}
\usepackage{dcolumn}
\usepackage{bm}
\newcommand{\ignore}[1]{}
\usepackage{epsfig}
\usepackage{color}

\begin{document}

\title{Temporal disorder in up-down symmetric systems}

\author{Ricardo Mart\'inez-Garc\'ia}
\affiliation{IFISC, Instituto de F\'isica Interdisciplinar y Sistemas Complejos (CSIC-UIB),
  E-07122 Palma de Mallorca, Spain}

\author{Federico Vazquez}
\affiliation{Max-Planck-Institut f\"ur
  Physik Komplexer Systeme, N\"othnitzer Str. 38, 01187 Dresden,
  Germany} 

\author{Crist\'obal L\'opez} 
\affiliation{IFISC, Instituto de F\'isica Interdisciplinar y Sistemas Complejos (CSIC-UIB),
  E-07122 Palma de Mallorca, Spain}

 \author{Miguel A. Mu\~noz}
\affiliation{Instituto Carlos I de F\'isica Te\'orica y Computacional, Universidad
de Granada, 18071 Granada, Spain}
\date{\today}
\begin{abstract}

  The effect of temporal disorder on systems with up-down Z2
  symmetry is studied. In particular, we analyze two well-known
  families of phase transitions: the Ising and the generalized voter
  universality classes, and scrutinize the consequences of placing
  them under fluctuating global conditions. We observe that
  variability of the control parameter induces in both classes
  ``Temporal Griffiths Phases'' (TGP).  These recently-uncovered
  phases are analogous to standard Griffiths Phases appearing in
  systems with quenched spatial disorder, but where the roles of space
  and time are exchanged. TGPs are characterized by broad regions in
  parameter space in which (i) mean first-passage times scale
  algebraically with system size, and (ii) the system response
  (e.g. susceptibility) diverges.   Our results confirm that TGPs are
  quite robust and ubiquitous in the presence of temporal disorder.
  Possible applications of our results to examples in ecology are
  discussed.

\end{abstract}


\maketitle

\section{Introduction}
\label{secintro}

Systems with up-down $Z_{2}$ symmetry --including the Ising model-- are
paradigmatic in Statistical Mechanics. Some of
them --such as the voter model-- exhibit absorbing states,
a distinctive feature of non-equilibrium dynamics
\cite{hinri,odor,GG,marro}. Absorbing states are configurations of the
system characterized by the lack of fluctuations, where the dynamics
becomes frozen and the system remains trapped. In the last years, a
great interest has been given to this class of models with two
symmetric absorbing states
\cite{hinri,odor,GG,marro,alhammal1,dornic,lipowski,droz,Vazquez-2008,Blythe},
which are of high relevance
in diverse problems in the ecological, biological, and social
sciences, such as species competition \cite{spcomp}, neutral theories
of biodiversity \cite{Durret}, allele frequency in genetics
\cite{allele}, opinion formation \cite{opinion}, epidemics propagation
\cite{Pinto}, or language spreading \cite{language}.

Phase transitions into absorbing states are quite universal. Systems
exhibiting one absorbing state belong generically to the very robust
Directed Percolation (DP) universality class and share the same set of
critical exponents, amplitude ratios, and scaling functions. When this
general rule is broken it is so owing to the presence of some
additional symmetry or conservation law \cite{hinri,odor,GG,marro}.  This
is the case of the class of systems with $Z_2$-symmetric absorbing
states, which may exhibit a phase transition with critical scaling
differing from DP, usually referred as Generalized voter (GV), also
called ``parity conserving'', ``DP2'' or ``directed Ising'',
universality class (see \cite{hinri,odor,dornic} and references
therein). Analytical and numerical studies
\cite{dornic,droz,alhammal1,Vazquez-2008,Blythe} have shown that,
depending on some details, $Z_2$-symmetric models may undergo either a
unique GV-like phase transition separating an active/symmetric phase
from an absorbing one or, alternatively, such a transition can split
into two separate ones: an Ising-like transition in which the $Z_2$
symmetry is broken, and a second DP-like transition below which the
broken-symmetry phase collapses into the corresponding absorbing
state.  In particular, a general stochastic theory, aimed at capturing
the phenomenology of these systems, was proposed in \cite{alhammal1};
depending on general features it may exhibit a DP, an Ising, or a GV
transition.

In many situations, $Z_{2}$ symmetric systems are not isolated but,
instead, affected by external conditions or by environmental
fluctuations. The question of how external variability affects
diversity, robustness, and evolution of complex systems, is of outmost
relevance in different contexts. Take, for instance, the example of
the neutral theory of biodiversity: if there are two $Z_2$-symmetric
(or neutral) species competing, what happens if depending on
environmental conditions one of the two species is favored at each
time step in a symmetric way?  Does such environmental variability
enhance species coexistence or does it hinder it?
\cite{Leigh,Chesson,Calabrese-2010,Vazquez-2010,dOdorico2009}.

Motivated by these questions, we study how basic properties of
$Z_2$ symmetric systems, such as response functions and
first-passage times, are affected by the presence of temporal
disorder.

Some previous works have explored the effects of fluctuating global
conditions in simple models exhibiting phase transitions
\cite{jensen,alonso,kamenev}. Temporal disorder has been shown
to be a highly relevant perturbation around DP phase transitions in
all dimensions (in apparent contradiction with  the Harris criterion for the
relevance of disorder \cite{jensen}), while temporal disorder has been
shown to be relevant at the Ising transition only at and above three
dimensions.  More recently, a modified version of the simplest
representative of the DP class --i.e. the Contact-Process-- equipped
with temporal disorder was studied in \cite{Vazquez-2011}. In this
model, the control parameter (birth probability) was taken to be a
random variable, varying at each time unit. As the control parameter
is allowed to take values above and below the transition point of the
pure contact process, the system alternates between the tendencies to
be active or absorbing. As shown in \cite{Vazquez-2011} this dynamical
frustration induces a logarithmic type of finite-size scaling at the
transition point and generates a subregion in the active phase
characterized by a generic algebraic scaling (rather than the usual
exponential, Kramers-like, behavior) of the extinction times with system size. More
strikingly, this subregion is also characterized by generic
divergences in the system susceptibility, a property which is reserved
for critical points in pure systems.  This phenomenology is akin to the
one in systems with quenched ``spatial'' disorder \cite{Vojta-2006},
which show algebraic relaxation of the order parameter, and
singularities in thermodynamic potentials in broad regions of
parameter space: the so-called, Griffiths Phases
\cite{griffiths}. The remarkable peculiarities of standard Griffiths
phases stem from the existence of (exponentially) rare --locally
ordered-- regions which take a (exponentially) long time to decay,
inducing an anomalously slow decay in the disordered phase.

In the case of temporal disorder, an analogy with Griffiths phases can
be made, in the sense that very long (exponentially rare) time
intervals (corresponding to an absorbing phase of the pure model) of
the control parameter have a large influence on the system dynamics
even when the overall system is in its active phase. These
phenomenological similarities between systems with spatial and
temporal disorder led us to introduce the concept of ``Temporal
Griffiths Phases'' (TGP) \cite{Vazquez-2011}.

In order to investigate whether the anomalous behavior that leads to
TGPs around absorbing state (DP) phase transitions is a universal
property of systems in other universality classes --and in particular,
in up-down symmetric systems-- we study the possibility of having TGPs
around Ising and GV transitions. We scrutinize simple models in these
two classes and assume that the corresponding control parameter
changes randomly in time, fluctuating around the transition point of
the corresponding pure model, and
study the susceptibility as well as mean-first passage times. We
mainly focus on the mean-field (high dimensional) limit, since it
allows for analytical treatment via a Langevin approach, but we also
provide numerical results and some theoretical considerations for low
dimensional systems.
    
The paper is organized as follows. In section \ref{MFmodels}, we
develop a general mean-field description of models with varying
control parameters in terms of collective variables. In section
\ref{secising} and \ref{secGV}, we show analytical and numerical
results for the Ising and GV transitions, respectively. In section
\ref{secsummary}, a short summary and conclusions are presented.

\section{Mean-field theory of $Z_2$-symmetric models with 
temporal disorder}
\label{MFmodels}

Interacting particle models evolve stochastically over time. A useful technique
to study such systems is the mean-field (MF) approach, which implicitly
assumes a ``well-mixed'' situation, where each particle can interact with any
other,  providing a sound approximation in high dimensional systems.
One way in which the mean-field limit can be seen at work is by analyzing
a fully connected network (FCN), where each node
(particle) is directly connected to any other else, mimicking an infinite
dimensional system.
 
In the models we study here, states can be labeled with occupation-number
variables $\rho_i$ taking a value $1$ if node $i$ is occupied or $0$
if it is empty, or alternatively by spin variables $S_{i}=2 \rho_i-1$,
with $S_{i}=\pm 1$. Using these latter, the natural order parameter is
the magnetization per spin, defined as
\begin{equation}
\label{magdef}
 m=\frac{1}{N}\sum_{i=1}^{N}S_{i},
\end{equation} 
where $N$ is the total number of particles in the system.  The master
equation for the probability $P(m,t)$ of having magnetization $m$ at a given time $t$, is 
\begin{eqnarray}
\label{prob} 
P(m,t+1/N) &=& \omega_{+}\left(m-2/N,b\right) 
P\left(m-2/N,t\right) \\
&+&\omega_{-}\left(m+2/N,b\right)P\left(m+2/N,t\right) \nonumber \\ 
&+&\left[1- \omega_{-}(m,b)-\omega_{+}(m,b)\right]P(m,t), \nonumber 
\end{eqnarray} 
where $\omega_{\pm}(m,b)$ are the transition probabilities from a
state with magnetization $m$ to a state with magnetization $m \pm
2/N$.  This describes a process in which a ``spin'' is randomly
selected at every time-step (of length $dt = 1/N$), and inverted with
a probability that depends on $m$ and the control parameter $b$.  The
allowed magnetization changes in an individual update, $\Delta m=\pm
2/N$, are infinitesimally small in the $N\rightarrow\infty$ limit.
In this limit, one can perform a standard Kramers-Moyal
expansion \cite{vanKampen,Gardiner} leading to the Fokker-Planck equation 
\begin{equation} 
\label{fokker} 
\frac{\partial P(m,t)}{\partial
t}=-\frac{\partial}{\partial m}\left[f(m,b)P(m,t)\right]
+\frac{1}{2}\frac{\partial^{2}}{\partial m^{2}}\left[g(m,b)P(m,t)\right],
\end{equation} 
with drift and diffusion terms given, respectively, by
\begin{eqnarray} 
f(m,b)&=&2\left[\omega_{+}(m,b)-\omega_{-}(m,b)\right],
\label{langedrift} \\ \nonumber \\
g(m,b)&=&\frac{4\left[\omega_{+}(m,b)+\omega_{-}(m,b)\right]}{N}. 
\label{langediff}
\end{eqnarray}
From Eq.~(\ref{fokker}), and working in the It\^{o} scheme (as
justified by the fact that it comes from a discrete in time equation
\cite{horsthemke}), its equivalent Langevin equation is
\cite{Gardiner}
\begin{equation}
\label{langevin}
 \dot{m}=f(m,b)+\sqrt{g(m,b)} \, \eta(t),
\end{equation}
where the dot stands for time derivative, and $\eta(t)$ is a Gaussian
white noise of zero-mean and correlations
$\langle\eta(t)\eta(t')\rangle=\delta(t-t')$. The diffusion term is
proportional to $1/\sqrt{N}$, and therefore, it vanishes in the
thermodynamic limit ($N \to \infty$), leading to a deterministic
equation for $m$.

\begin{figure}
\centering 
\includegraphics[width=0.45\textwidth]{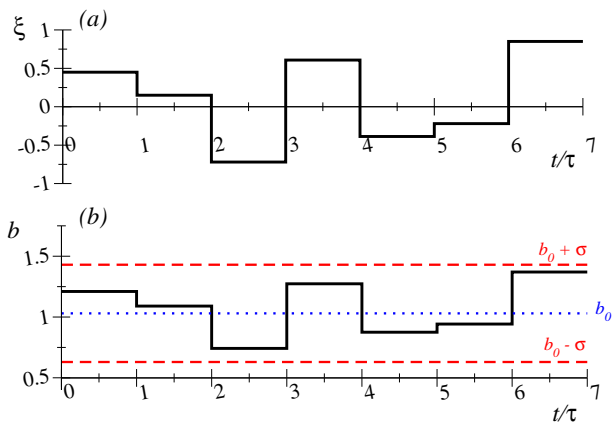}
\caption{(Color online). (a) Typical realization of the colored noise $\xi(t)$, a step
  like function that takes values between $+1$ and $-1$. (b)
  Stochastic control parameter $b(t)=b_{0}+\sigma\xi(t)$ according to
  the values of the noise in (a), $b_{0}=1.03$ and $\sigma=0.4$.}
\label{noise}
\end{figure}

\begin{figure}
\centering 
\includegraphics[width=0.35\textwidth]{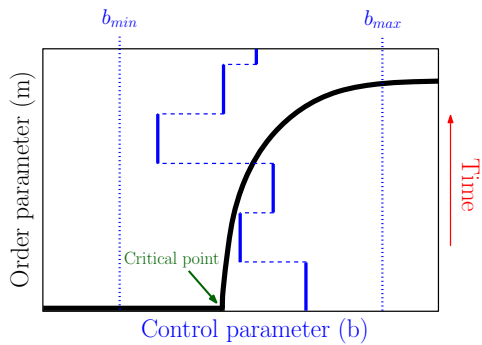}
\caption{(Color online). Schematic representation of the fluctuating control parameter
  in the Ising model with Glauber dynamics. It makes the system shift
  between the ordered phase to the disordered one.}
\label{noiseising}
\end{figure}

The drift and diffusion coefficients in Eq.~(\ref{langevin}) depend
not only on the magnetization, but also on the parameter $b$. To
analyze the behavior of the system when $b$ changes randomly over
time, and following previous works \cite{Vazquez-2010,Vazquez-2011},
we allow $b$ to take a new random value, extracted from a uniform
distribution, in the interval $(b_0-\sigma,b_0+\sigma)$ at each MC
step, i.e., every time interval $\tau = 1$.
Thus, we assume that the dynamics
of $b(t)$ obeys an Ornstein-Uhlenbeck process \begin{equation}
 b(t)=b_{0}+\sigma \, \xi(t),
\label{controlrandom}
\end{equation}
where $\xi(t)$ is a step-like function that randomly fluctuates
between $-1$ and $1$, as depicted in Fig.~\ref{noise}a. Its average correlation is
\begin{eqnarray}
\label{corr}
 \overline{<\xi(t)\xi(t+\Delta t)>} = \left\{ \begin{array}{ll}
\frac{1}{3}(1-|\Delta t|/\tau) & \mbox{for $|\Delta t|<\tau$}\\ \\
0 & \mbox{for $|\Delta t| >\tau$,}
\end{array}
\right.
\end{eqnarray}
where the bar stands for time averaging.  The parameters $b_0$ and
$\sigma$ are chosen with the requirement that $b$ takes values at both
sides of the transition point of the \emph{pure model} (see
Fig.~\ref{noise}b), that is, the model with constant $b$.  Thus, the
system randomly shifts between the tendencies to be in one phase or
the other (see Fig.~\ref{noiseising}).
The model presents both \emph{intrinsic} and \emph{extrinsic}
fluctuations, as represented by the white noise $\eta(t)$ and the
colored noise $\xi(t)$, respectively.  Plugging the expression Eq.~
(\ref{controlrandom}) for $b(t)$ into Eq.~(\ref{langevin}), and
retaining only linear terms in the noise one readily obtains
\begin{equation} 
\label{2nois}
\dot{m}=f_{0}(m)+\sqrt{g_{0}(m)} \, \eta(t)+j_{0}(m) \, \xi(t),
\end{equation}
where $f_{0}(m) \equiv f(m,b_0)$, $g_{0}(m) \equiv g(m,b_0)$ and
$j_{0}(m)$ is a function determined by the functional form of
$f(m,b)$, that might also depend on $b_{0}$.  To simplify the
analysis, we assume that relaxation times are much longer than the
autocorrelation time $\tau$, and thus take the limit $\tau \to 0$ in
the correlation function Eq.~(\ref{corr}), and transform the external
colored noise $\xi$ into a Gaussian white noise with effective
amplitude $K
\equiv\int_{-\infty}^{+\infty}\overline{<\xi(t)\xi(t+\Delta t)>}
\,d\Delta t = \tau/3$.  Then, we combine the two white noises into an
effective Gaussian white noise, whose square amplitude is the sum of
the squared amplitudes of both noises \cite{Gardiner}, and finally
arrive at
\begin{equation}
\label{finallange}
\dot{m}=f_{0}(m)+\sqrt{g_{0}(m)+Kj_{0}^{2}(m)} \, \gamma(t),
\end{equation}
where $\langle \gamma(t) \rangle=0$ and $\langle \gamma(t) \gamma(t')
\rangle=\delta(t-t')$.

In the next two sections, we analyze the dynamics of the kinetic Ising
model with Glauber dynamics and a variation of the voter model (the,
so-called, q-voter model) --which are representative of the Ising and
GV transitions respectively-- in the presence of external noise. For
that we follow the strategy developed in this section to derive
mean-field Langevin equations and present also results of numerical
simulations (for both finite and infinite dimensional systems), as
well as analytical calculations.

\section{Ising transition with temporal disorder}
\label{secising}

We consider the kinetic Ising model with Glauber dynamics
\cite{Glauber-1969}, as defined by the following transition rates
\begin{equation}\label{transgl}
 \Omega_{i}(S_{i}\rightarrow-S_{i})=\frac{1}{2}\left[1-S_{i}\,
{\rm \tanh}\left(\frac{b}{2d}\sum_{j\in \langle i \rangle}S_{j}\right)\right].
\end{equation}
The sum extends over the $2d$ nearest neighbors of a given spin $i$ on
a $d$-dimensional hypercubic lattice, and $b=J\beta$ is the control
parameter. $J$ is the coupling constant between spins, which we set to
$1$ from now on, and $\beta=(k_{B}T)^{-1}$. Note that $b$ in this case
is proportional to the inverse temperature.

\subsection{The Langevin equation}
\label{Lang-Eq}

In the mean-field case, the cubic lattice is replaced by a fully-connected network in
which the number of neighbors $2d$ of a given site is simply $N-1$.  Then, the 
transition rates of Eq.~(\ref{transgl}) can be expressed as
\begin{equation}
\label{transglmf}
 \Omega_{\pm}(m,b) \equiv \Omega(\mp \to \pm) = 
\frac{1}{2}\left[1\pm{\rm tanh}\left(b \, m\right)\right].
\end{equation}
which implies 
$ \omega_{\pm}(m,b)=\frac{1 \mp m}{2} \, \Omega_{\pm}(m,b)$
for jumps in the magnetization. Following the steps in the previous
section, and expanding $\Omega_{\pm}$ to third order in $m$, we obtain
\begin{equation} \label{isinglangevin}
 \dot{m}=a_{0}m-c_{0}m^{3}+\sqrt{\frac{1-b_{0} m^{2}}{N}+
K\sigma^{2}m^{2}(1-b_{0}^{2}m^{2})^{2}}\,\,\gamma(t),
\end{equation}
where $b_{0}$ is the mean value of the stochastic control parameter,
$a_{0}\equiv b_{0}-1$, and $c_{0}\equiv b_{0}^{3}/3$.

The potential $V(m) = -\frac{a_{0}}{2} m^2 +\frac{c_{0}}{4}m^{4}$
associated with the deterministic term of Eq.~(\ref{isinglangevin})
has the standard shape of the Ising class, that is, of systems
exhibiting a spontaneous breaking of the $Z_2$ symmetry. A single
minimum at $m=0$ exists in the disordered phase, while two symmetric
ones, at $\pm\sqrt{a/c}$ exist below the critical point.

\subsection{Numerical Results}

\begin{figure} \centering
  \includegraphics[width=0.4\textwidth]{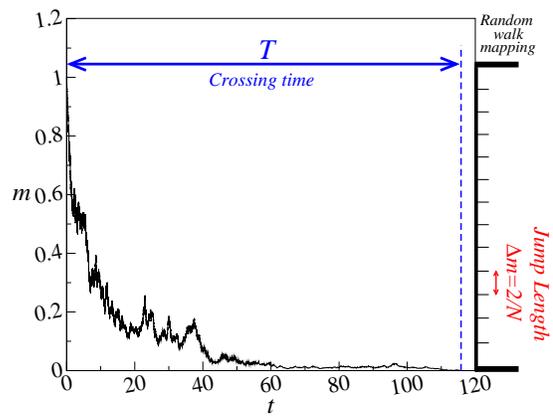} \caption{(Color online). 
    Single realization of the stochastic process. The system starts with all
    the spins in the same state $(m=1)$ and the dynamics is stopped
    when it crosses $m=0$, which defines the crossing time in the
    Ising model. We take $\sigma=0.4$, $b_{0}=0.98$ and system size
    $N=10^{6}$. On the right margin we sketch the mapping of the
    problem to a Random Walk with jump length $|\Delta m|=2/N$.}
\label{single}
\end{figure}

\begin{figure} \centering
  \includegraphics[width=0.4\textwidth]{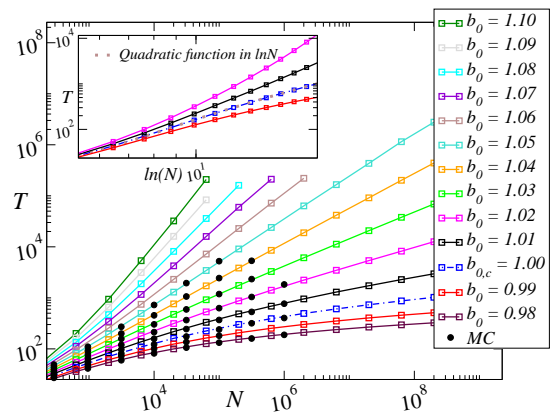} \caption{(Color online). 
    Main:
    Log-log plot of the crossing time $T(N)$ for the Ising Model with
    Glauber dynamics in mean field. Values of the control parameter
    from $b_{0}=0.98$ (bottom) to $b_{0}=1.10$ (top) are shown.
    Monte Carlo simulations on a FCN
    (circles) and numerical integration of the Langevin equation Eq.~
    (\ref{isinglangevin}) with $\sigma=0.4$ (squares and interpolation 
    with solid lines). There is a
    region with generic algebraic scaling of $T(N)$ and continuously
    varying exponents, $b_{0}\in[1.01,1.10]$. Inset: log-log plot 
    of $T(N)$ vs. $\ln N$.
    At criticality (dotted-dashed line) the scaling is fitted to a 
    quadratic function in $\ln N$.}
\label{passtisingmf}
\end{figure}

In this section we study two magnitudes that were shown to be relevant
in systems with temporal disorder \cite{Vazquez-2011}: the mean
\emph{crossing time} (or mean-first passage time) and the
susceptibility.  The crossing time is the time employed by the system
to reach the disordered zero-magnetization state for the first time,
starting from a fully ordered state with $|m|=1$ (see
Fig.~\ref{single}).  Crossing times were calculated by numerically
integrating Eq.~(\ref{isinglangevin}) for different realizations of
the noise $\gamma$ and averaging over many independent realizations.
These integrations were performed using a standard stochastic
Runge-Kutta scheme (note that, the noise term does not have any
pathological behavior at $m=0$ as occurs in systems with absorbing
states, for which more refined integration techniques are required
\cite{Dornic}) .  Results are shown in Fig.~\ref{passtisingmf}.


To estimate the critical point, we calculated the time evolution of
the average magnetization $\langle m \rangle(t)$ by integrating the
Langevin equation Eq.~(\ref{isinglangevin}), and also by performing
Monte Carlo simulations of the particle system on a fully connected
network.  At the critical point $b_{0,c}$ the magnetization decays to
zero as $\langle m \rangle \sim t^{-\beta}$.  We have estimated
$b_{0,c} = 1$, which coincides with the pure case critical point
$b_{c,pure}=1$: the critical point in the presence of disorder in
mean-field is not shifted with respect to the pure system, in
agreement with the analytical calculation in appendix \ref{App-A}.  At
this critical point, as it is characteristic of TGPs
\cite{Vazquez-2011}, a scaling of the form $T \sim [\ln N]^{\alpha}$
is expected.  The numerically determined exponent value $\alpha \simeq
2.81$ for $\sigma=0.4$ is higher than the exponent $\alpha=2$ of the
asymptotic analytical prediction Eq.~(\ref{apA:final}), probably
because of the asymptotic regime in $\ln N$ has not been reached.
Instead, the behavior for arbitrary values of $N$ appears to be a
second order polynomial in $\ln N$, as we can see in
Eq.~(\ref{quadraticpol}).  Indeed, the numerical data is well fitted
by the quadratic function $a\,(\ln N)^2 + b \, \ln N + c $ (see inset
of Fig.~\ref{passtisingmf}).This is to be compared with the standard
power-law scaling $T\sim N^{\beta}$ characteristic of pure systems,
i.e. for $\sigma=0$. Moreover, a broad region showing algebraic
scaling $T \sim N^{\delta}$ with a continuously varying exponent
$\delta(b_0)$ ($\delta \rightarrow 0$ as $b_{0} \rightarrow
b^{+}_{0,c}$) appears in the ordered phase $b_0 > b_{0,c}$. Both
$\alpha$ and $\delta$ are not universal and depend on the noise
strength $\sigma$. Finally, in the disordered phase the scaling of $T$
is observed to be logarithmic, $T \sim \ln N$.

\begin{figure} \centering
  \includegraphics[width=0.4\textwidth]{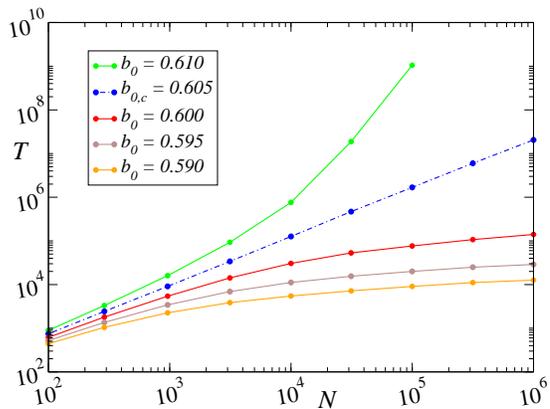} \caption{(Color online).
    Log-log plot of the crossing time $T(N)$ for the Ising Model with
    Glauber dynamics in $d=2$. Values of the control parameter
    from $b_{0}=0.590$ (bottom) to $b_{0}=0.610$ (top) are shown.
    Monte Carlo simulations on a regular
    cubic lattice with $\sigma=0.4$ (lines are interpolations).
    We observe a power law scaling
    at the critical point (dotted-dashed line). TGP are not observed, 
    crossing time scales exponentially in the ordered phase (light green, upper, line).}
  \label{escape2d} \end{figure}

\begin{figure} \centering
  \includegraphics[width=0.4\textwidth]{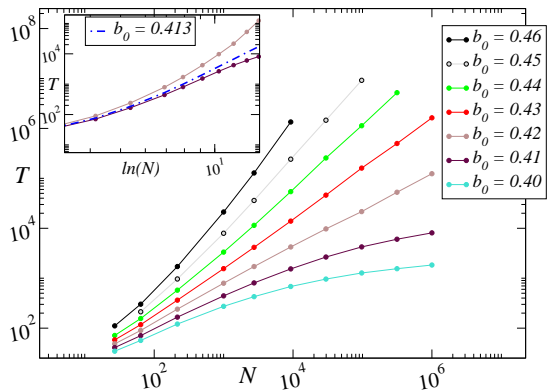} \caption{(Color online). Main:
    Log-log plot of the crossing time $T(N)$ for the Ising Model with
    Glauber dynamics in $d=3$. Monte Carlo simulations on a regular
    cubic lattice with $\sigma=0.4$ (lines are interpolations).
    Values of the control parameter
    from $b_{0}=0.40$ (bottom) to $b_{0}=0.46$ (top) are shown.
    There is a region $b \in [0.42,
    0.46]$ with generic algebraic scaling of $T(N)$ and continuously
    varying exponents. Inset: log-log plot of $T(N)$ vs. $\ln (N)$. It
    is estimated at criticality (dotted-dashed line) $T \sim (\ln N)^{5.29}$.}
  \label{passtising3D} \end{figure}

We have also performed Monte Carlo simulations of the time-disordered
Glauber model on two- and three-dimensional cubic lattices with
nearest neighbor interactions.  The critical point was computed
following standard methods, that is, by looking for a power law
scaling of $\langle m \rangle $ versus time, as we mentioned above.
In $d=2$, a shift in the critical point was found: from
$b_{c,pure}=0.441(1)$ in the pure model to $b_{0,c}=0.605(1)$ for
$\sigma=0.4$. However, the scaling behavior of $T$ with $N$ resembles
that of the pure model, with $T\sim N^{\beta}$ at criticality (with an
exponent numerically close to that of the pure model \cite{marro}),
and an exponential growth $T \sim \exp(c N)$, where $c$ is a positive
constant, in the ordered phase (Arrhenius law) ) { (see
  Fig.~\ref{escape2d})}.  Thus, no region of generic algebraic scaling
appears in this low-dimensional system. On the contrary, in $d=3$,
results qualitatively similar to mean-field ones are recovered (see
Fig.~\ref{passtising3D}). The critical point is shifted from
$b_{c,pure}=0.222(1)$ (calculated in \cite{heuer}) to
$b_{0,c}=0.413(2)$, with a critical exponent $\alpha(d=3)=5.29$ for
$\sigma=0.4$, and generic algebraic scaling in the ordered phase. In
conclusion, our numerical studies suggest that the lower critical
dimension for the TGPs in the Ising transition is $d_c=3$. This is in
agreement with the analytical finding in \cite{alonso}, establishing
that temporal disorder is irrelevant in Ising-like systems below three
dimensions. This result is to be compared with $d_c=2$ numerically
reported for the existence of TGPs in DP-like transitions
\cite{Vazquez-2011} (observe, however, that temporal disorder, in this
case, affects the value of critical exponents at criticality in all
spatial dimensions). Further studies are needed to clarify the
relation between disorder-relevance at criticality and the existence
or not of TGPs.

We have also measured the susceptibility $\chi$, defined as the response
function to an external field $h$ in the vanishing field limit 
\begin{equation} 
\chi=\lim_{h\rightarrow 0} \frac{\partial \langle m \rangle }{\partial h},
\end{equation} 
where $\langle m \rangle$ denotes the stationary magnetization averaged over many
independent realizations. In the presence of an external field, the
transition rates become
 $\Omega_{\pm}(m,b) = \frac{1}{2}\left[1\pm{\rm tanh}\left(b \, m+h\right)\right]$.
 Expanding the hyperbolic tangent up to third order in $m$ and to first order
 in $h$, we obtain the following  Langevin equation
 \begin{equation} \label{ising-langevin-h} \dot{m}=a_{0}m-c_{0}m^{3}+
   h\left(1-b^{2}m^{2}\right) + \sqrt{K} \sigma m
   (1-b_{0}^{2}m^{2})\,\,\gamma(t),
\end{equation}
where we have considered the $N \to \infty$ limit $(g_{0}=0)$.

\begin{figure}
  \centering \includegraphics[width=0.4\textwidth]{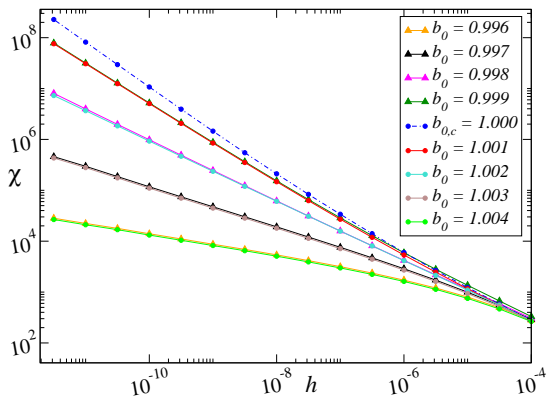}
  \caption{(Color online). Main: Log-log plot of the susceptibility as a function of
    the external field for different values of $b_0 \in
    [b_{0,c}-\sigma^{2}/2, b_{0,c}+\sigma^{2}/2]$ with $\sqrt{K}\sigma=0.1$,
    obtained by integrating Eq.~(\ref{ising-langevin-h}). Generic
    divergences with symmetric exponents around the critical value
    $b_{0.c}=1$ (dotted-dashed line) are observed.} \label{figsus} \end{figure}

The average magnetization $\langle m \rangle $ for a given field $h$ was
calculated by integrating the Langevin equation and then taking
averages over noise realizations. The susceptibility can be computed,
for different values of $b_0$, as the derivative of $\langle m \rangle$ with
respect to $h$. Generic divergences of the form $\chi \sim
h^{\upsilon} + \mbox{Constant}$ (with $\upsilon <0$ as $h \to 0$)
appear in a broad region $b_{0} \in
[b_{0,c}-\sigma^{2}/2,b_{0,c}+\sigma^{2}/2]$, centered around
$b_{0,c}$, with symmetric exponents around the critical point (see
Fig.~\ref{figsus}). These results agree with those obtained through
Monte Carlo simulations on a FCN (not shown). In finite dimensions, given the
required large systems sizes  and small fields,  we
could not conclude about the existence or not of generic divergences.

\subsection{Analytical results} 
\label{isinganal}

Let us consider the Langevin equation Eq.~(\ref{isinglangevin}) in the
thermodynamic limit ($g_{0}(m)=0$). Given that the remaining intrinsic
noise comes from a transformation of a colored noise into a white
noise, the Stratonovich interpretation is to be used to obtain its
associated Fokker-Planck equation (see e.g.  \cite{horsthemke})
\begin{eqnarray} \label{fpst} \frac{\partial P(m,t)}{\partial
    t}&=&-\frac{\partial}{\partial m} \Bigg\{ \left[
    f_{0}(m)+\frac{K}{2}j_{0}(m)j'_{0}(m)\right] P(m,t)\Bigg\}
  \nonumber \\
  &+&\frac{1}{2}\frac{\partial^{2}}{\partial m^{2}} \Big\{ K
  j^{2}_{0}(m)P(m,t)\Big\}. \end{eqnarray} 
Imposing the detailed
balance (fluxless) condition, it is straightforward to obtain the
steady state solution
\begin{equation}
 P_{st}(m)\propto\exp\left(-\frac{c_{0}m^{2}}{K\sigma^{2}}\right)
|m|^{\frac{2 a_0}{K\sigma^2} -1},
\end{equation}
with a power-law singularity at the origin; this is a distinctive
trait of a Langevin equation with linear multiplicative noise
\cite{grinstein,renmunoz}.  By performing a calculation analogous to
that in \cite{Vazquez-2011}, we have analytically computed the system
susceptibility and found that $\chi \sim h^{\upsilon} +
\mbox{Constant}$, as mentioned earlier, and in agreement with previous
results found in \cite{Vazquez-2011,grinstein,renmunoz}, with
\begin{equation}
 \upsilon=\frac{2 (b_{0}-1)}{K\sigma^2} -1.
\end{equation}
This, in particular, implies that the susceptibility diverges when
$\upsilon<0$ as $h \to 0$ or, in terms of the control parameter
$b_0=1+a_0$, when $b_0$ takes a value in the region
$1-\sigma^2/2<b_0<1+\sigma^2/2$ centered at the critical point
$b_{0,c}=1$. The values of the exponent $\upsilon$ agree well with
those of Fig.~\ref{figsus} at some distance from the critical point.
For instance, an analytical value $\upsilon =-0.40$ for $b_{0}=1.003$
corresponds to a numerical value $\upsilon_{num}=-0.39$, and
$\upsilon=-0.60$ for $b_{0}=-1.002$ to a value $\upsilon_{num}=-0.59$.
However, the analytical exponent $\upsilon = -1$ at the critical point
is not in good agreement with the numerical result
$\upsilon_{num}=-0.88$, indicating that the asymptotic regime has not
been numerically reached.

We next provide analytical results for the crossing time. Starting
from the N-independent Fokker-Planck equation Eq.~(\ref{fpst}), an
effective dependence on $N$ is implemented by calculating the
first-passage time to the state $m=|2/N|$ rather than $m=0$. This is
equivalent to the assumption that the system reaches the zero
magnetization state with an equal number $N_+=N_-=N/2$ of up and down
spins when $|m|<2/N$, that is, when $N/2-1<N_+<N/2+1$. The mean-first
passage time $T$ associated with the Fokker-Planck
equation Eq.~(\ref{fpst}) obeys the differential equation \cite{Gardiner}
\begin{equation}\label{meanesc}
 \frac{K}{2}j^{2}_{0}(m)T''(m)+\left[ f_{0}(m)+\frac{K}{2}j_{0}(m)j'_{0}(m)\right]T'(m)=-1,
\end{equation}
with absorbing and reflecting boundaries at $|m|=2/N$ and $|m|=1$,
respectively.  The solution, starting at time $t=0$ from $m=1$ is given by 
\begin{equation}
\label{time}
T(m=1)=
2\int_{2/N}^{1}\frac{dy}{\psi(y)}\int_{y}^{1}\frac{\psi(z)}{Kj_{0}^{2}(z)}dz,
\end{equation}
where
\begin{equation}\label{psi1}
\psi(x)=\exp \Bigg\{ \int_{2/N}^{x} \frac{2f_{0}(x')+
Kj_{0}(x')j'_{0}(x')}{Kj_{0}^{2}(x')} dx' \Bigg\}. 
\end{equation}
Computing these integrals (see Appendix \ref{App-A}) we obtain
\begin{eqnarray} 
\label{timeana}
T \sim \left\{ \begin{array}{ll}
{\ln N}/(b_{0}-1) & \mbox{for $b_{0} < 1$} \\
3(\ln N)^{2} /\sigma^{2} & \mbox{for $b_{0} = 1$} \\
N^{\frac{6(b_{0}-1)}{\sigma^{2}}} & \mbox{for $b_{0} > 1$.}
\end{array}
\right.
\end{eqnarray}
These expressions qualitatively agree with the numerical results of
Fig.~\ref{passtisingmf}, showing that $T$ grows logarithmically with
$N$ in the absorbing phase $b_0<1$, as a power law in the active phase
$b_0>1$, and as a power of $\ln N$ (i.e. poly-logarithmically) at the
transition point $b_{0,c}=1$. The exponents $\delta=6(b_{0}-1)/\sigma^{2}$ 
do not agree well
with the numerically determined exponents. This is probably due to to
the fact that we have neglected the $1/\sqrt{N}$ term by taking
$g_0=0$, which becomes of the same magnitude as the $j_0$ term when
$|m|$ approaches $2/N$. Indeed, this was confirmed (not shown) by
testing that analytical expressions Eq.~(\ref{timeana}) agree very
well with numerical integrations of Eq.~(\ref{isinglangevin})
performed for $g_{0}=0$, and setting the crossing point at $m=2/N$. In
summary, this analytical approach reproduces qualitatively --and in some
cases quantitatively-- the above reported non-trivial phenomenology.


\section{Generalized Voter transition with temporal disorder}
\label{secGV}

We study in this section the GV transition \cite{dornic}, which
appears when a $Z_2$-symmetry system simultaneously breaks the
symmetry and reaches one of the two absorbing states. A model
presenting this type of transition is the nonlinear $q$-voter model,
introduced in \cite{qvoter}. The microscopic dynamics of this
nonlinear version of the voter model consists in randomly picking a
spin $S_i$ and flipping it with a probability that depends on the
state of $q$ randomly chosen neighbors of $S_i$ (with possible
repetitions). If all neighbors are at the same state, then $S_i$
adopts it with probability $1$ (which implies, in particular, that the two
completely ordered configurations are absorbing). Otherwise, $S_i$
flips with a state-dependent probability
 \begin{equation}
 f(x,b)=x^{q}+b[1-x^{q}-(1-x)^{q}],
\label{probqvoter}
\end{equation}
where $x$ is the fraction of disagreeing (antiparallel) neighbors and
$b$ is a control parameter. Three types of transitions, Ising, DP and
GV can be observed in this model depending on the value of $q$
\cite{qvoter}. Here, we focus on the $q=3$ case, for which a unique GV
transition at $b_c=1/3$ has been reported \cite{qvoter}.

\subsection{The Langevin equation}

\begin{figure} \centering
  \includegraphics[width=0.20\textwidth]{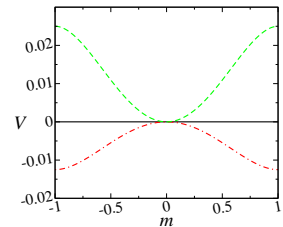}
  \caption{(Color online). Potential for the GV transition in a mean field approach.
    The dashed, solid and dot-dashed lines correspond to the
    paramagnetic phase, critical point, and the ferromagnetic phase,
    respectively.}
\label{potential-GV}
\end{figure}

In the MF limit (FCN) \cite{mf}, the fractions of antiparallel
neighbors of the two types of spins $S_i =1$ and $S_i=-1$ are
$x=(1-m)/2$ and $x=(1+m)/2$, respectively. Thus, the transition
probabilities are \begin{equation} \label{w+-}
 \omega_{\pm}(m,b)=\frac{1\mp m}{2}\,f\left(\frac{1\pm m}{2},b\right).
\end{equation}
Following the same steps as in the previous section, we obtain the Langevin equation
\begin{eqnarray}
\label{GVlangevin}
&&\dot{m}=\frac{1-3b_{0}}{2}m(1-m^{2})+ \\
&&\sqrt{\frac{\left(1-m^{2}\right)(1+6b_{0}+m^{2})}{N}+
  \frac{9 K}{4}\sigma^{2}m^{2}(1-m^{2})^{2}}\,\, \gamma(t). \nonumber
\end{eqnarray}
Let us remark that the potential in the nonlinear voter model
(Fig.~\ref{potential-GV}) differs from that for the Ising model. Owing
to the fact that the coefficients of the linear and cubic term in the
deterministic part of Eq.~(\ref{GVlangevin}) coincide (except for
their sign), the system exhibits a discontinuous jump at the
transition point, where the potential minimum changes directly from
$m=0$ in the disordered phase to $m=\pm1$ in the ordered one.
Furthermore, the potential vanishes at the critical point
\cite{alhammal1}.

\subsection{Numerical Results}

\begin{figure}
\centering 
\includegraphics[width=0.4\textwidth]{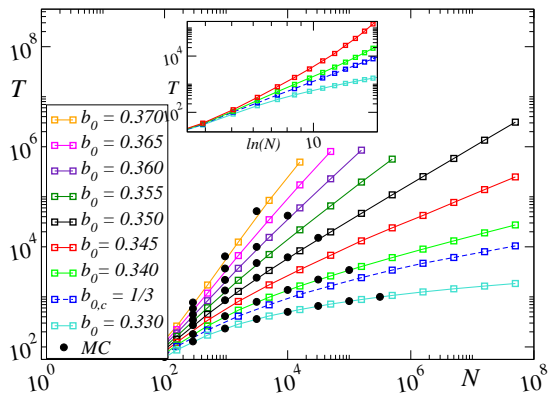}
\caption{(Color online). Main: Log-log plot of the ordering time as a function of the
  system size $N$ in the MF q-voter model. Monte Carlo simulations on a FCN
  (dots) and numerical integration of the Langevin equation Eq.~
  (\ref{GVlangevin}) for values of $b$ going from $0.330$ (bottom) to
  $0.370$ (top), and $\sigma=0.3$ (squares and solid lines interpolation). In the active phase a
  finite region with power law scaling is observed, $b_{0}\in[0.340,0.370]$. Inset: log-log
  plot of $T$ as a function of $\ln N$. At the critical point (dashed line) is $T
  \sim [\ln N]^{3.68}$.}
\label{timegv}
\end{figure}

The \emph{ordering time}, defined as the averaged time required to
reach a completely ordered configuration (absorbing state) starting
from a disordered configuration, is the equivalent of the crossing
time above. We have measured the mean ordering time $T$ by both,
integrating the Langevin equation Eq.~(\ref{GVlangevin}) and running Monte
Carlo simulations of the microscopic dynamics on FCNs and finite
dimensions. In Fig.~\ref{timegv} we show the MF results. We observe
that $T$ has a similar behavior to the one found for the mean crossing
time in the Ising model, and for the mean extinction time for the
contact process \cite{Vazquez-2011}. That is, a critical scaling $T
\sim [\ln N]^{\alpha}$ at the transition point $b_{0,c}=1/3$, with a
critical exponent $\alpha=3.68$ for $\sigma=0.3$, a logarithmic
scaling $T \sim \ln N$ in the absorbing phase $b_0<b_{0,c}$, and a
power law scaling $T \sim N^{\delta}$ with continuously varying
exponent $\delta(b_0)$ in the active phase $b_0>b_{0,c}$.

Monte Carlo simulations on regular lattices of dimensions $d=2$ and
$d=3$ revealed that there is no significant change in the scaling
behavior respect to the pure model (not shown).  The critical point
shifts in $d=2$ and remains very close to its mean-field value in
$d=3$, but results are compatible with the usual critical (pure) voter scaling
$T_{2d}\sim N\ln N$ and $T_{3d}\sim N$.  In the absorbing
phase $T$ grows logarithmically with $N$, while in the active phase
$T$ grows exponentially fast with $N$, as in the pure-model case.
Therefore, in these finite dimensional systems we do not find any
TGP nor other anomalous effects induced by temporal disorder, although
we cannot numerically exclude their existence in $d=3$. Such
effects should be observable, only in higher dimensional systems
(closer to the mean-field limit).

\subsection{Analytical results}

The ordering time $T$ can be estimated by assuming that the dynamics
is described by the Langevin equation Eq.~(\ref{GVlangevin}), and
calculating the mean first-passage time from $m=0$ to any of the two
barriers located at $|m|=1$.  It turns out useful to consider the
density of up spins rather than the magnetization
\begin{equation} \label{change-rho}
 \rho \equiv \frac{1+m}{2}.
\end{equation}
 $T$ is the mean first-passage time to $\rho=0$
starting from $\rho=1/2$.  The Langevin equation for $\rho$ is
obtained from Eq.~(\ref{GVlangevin}), by neglecting the $1/\sqrt{N}$
term and applying the ordinary transformation of variables (which is done
 employing standard algebra, given that Eq.~(\ref{GVlangevin})
is interpreted in the Stratonovich sense) is
\begin{equation}
 \dot{\rho}=A(\rho)+\sqrt{K}C(\rho)\gamma(t),
\end{equation} 
with
\begin{eqnarray} \label{change-function}
 A(\rho)&=&a_{0}\rho(2\rho-1)(1-\rho), \nonumber \\
 C(\rho)&=&3\sigma\rho(2\rho-1)(1-\rho),
\end{eqnarray}
where $a_{0}=1-3b_{0}$.\\
Now, we can follow the same steps as in section \ref{isinganal} for
the Ising model, and find the equation for the mean first-passage time
$T(\rho)$ by means of the Fokker-Planck equation.  The solution is
given by (see Appendix B)
\begin{eqnarray}
\label{timeanagv}
T \sim \left\{ \begin{array}{ll}
{\ln N}/{\left(3b_{0}-1\right)} & \mbox{for $b_{0} < 1/3$} \\
{(\ln N)^{2}}/{3\sigma^{2}} & \mbox{for $b_{0} = 1/3$} \\
N^{\frac{2(b_{0}-1/3)}{\sigma^{2}}} & \mbox{for $b_{0} > 1/3$.}
\end{array}
\right.
\end{eqnarray}
These scalings, which qualitatively agree with the numerical results
of Fig.~\ref{timegv} for the $q$-voter, show that the behavior of $T$
is analogous to the one observed in the Ising transition of section
\ref{secising} and in the DP transition found in \cite{Vazquez-2011}.
Therefore, we conclude that TGPs appear around GV transitions in the
presence of external varying parameters in high dimensional systems.

 For the GV universality class the renormalization group fixed point
 is a non-perturbative one \cite{Canet}, becoming relevant in a
 dimension between one and two. A field theoretical implementation of
temporal disorder in this theory is still missing, hence, theoretical
predictions and sound criteria for disorder relevance are
not available.

\section{Summary and conclusions}
\label{secsummary}

We have investigated the effect of temporal disorder on phase
transitions exhibited by $Z_2$ symmetric systems: the (continuous)
Ising and (discontinuous) GV transitions which appear in many
different scenarios. We have explored whether temporal disorder
induces Temporal Griffiths Phases
as it was previously found in standard (DP) systems
with one absorbing state. By performing mean-field analyses as well as
extensive computer simulations (in both fully connected networks and
 in finite dimensional lattices) we found that TGPs can exist
around equilibrium (Ising) transitions (above $d=2$) and around
discontinuous (GV) non-equilibrium transitions (only in
high-dimensional systems).  Therefore, we confirm that TGPs may also
appear in systems with two symmetric absorbing states, illustrating
the generality of the underlying mechanism: the appearance of a
region, induced by temporal stochasticity of the control parameter,
where first-passage times scale as power laws of the system size and
where the susceptibility diverges.  Temporal disorder, makes the
ordered/active phase less stable and makes the system highly
susceptible to perturbations.  This appears to be a rather general and
robust phenomenon.

It also seems to be a general property that TGPs do not appear in low
dimensional systems, where standard fluctuations dominate over
temporal disorder. In all the cases studied so far, a critical
dimension $d_c$ --at and below which TGPs do not appear-- exist
($d_c=1$ for DP transitions, $d_c=2$ for Ising like systems, and $d_c
\simeq 3$ for GV ones).  Calculating analytically such a critical
dimension and comparing it with the standard critical dimension for
the relevance/irrelevance of temporal disorder at the critical point
(i.e. at the renormalization group non-trivial fixed point of the
corresponding field theory) remains an open and challenging task.

A relevant application of our results is found in models of ecosystems.
In this case, first-passage times are related to typical extinction
times, and studying how such extinction times are affected by system
size (e.g. habitat fragmentation) is a problem of outmost relevance.
Future research might be oriented to the effect of temporal disorder
 on the formation and dynamics of spatial structures.


\acknowledgments
R.M-G. is supported by the JAEPredoc program of CSIC. 
R.M-G. and C.L. acknowledge support 
from MICINN
(Spain) and FEDER (EU) through Grant No. FIS2007-
60327 FISICOS. MAM acknowledges financial support from the Spanish
MICINN-FEDER under project FIS2009-08451 and from Junta de Andaluc{\'\i}a
Proyecto de Excelencia P09FQM-4682. We are grateful to J.A. Bonachela
for useful discussions and a critical reading of the manuscript.


\appendix 
\section{Analytical calculations of the crossing time for the mean
  field Ising model}
\label{App-A}

The mean first passage time to reach an absorbing barrier at $|m|=2/N$
starting from $|m|=1$ can be expressed as \cite{Gardiner},
\begin{equation}\label{esctime}
T(m=1)=
2\int_{2/N}^{m=1}\frac{dy}{\psi(y)}\int_{y}^{1}\frac{\psi(z)}{Kj_{0}^{2}(z)}dz,
\end{equation}
with
\begin{equation}\label{psi}
\psi(z)=
\exp\int_{2/N}^{z}dz'\frac{2f_{0}(z')+Kj_{0}(z')j_{0}'(z')}{Kj_{0}^{2}(z')},
\end{equation}
which involves $6^{th}$ and $4^{th}$ order polynomial functions.
In order to make the integral simpler, functions are expanded up to $3^{rd}$
order, 
\begin{eqnarray}
f_0(m)+\frac{K}{2}j_{0}(m)j_{0}'(m)&\approx&m(r-sm^{2}) \nonumber \\
Kj_{0}^{2}(m)&\approx&\omega m^{2},
\end{eqnarray}
with $\omega \equiv \sigma^{2}/3$, $r \equiv a_{0}+\omega/2$, $s\equiv
(c_{0}+2\omega b_{0}^{2})$.  A second simplifying assumption is to
take $1$ as the lower integration limit in Eq.~(\ref{psi}) instead of $2/N$ (justified 
because $\psi(z)$ appears both in the numerator and the denominator
of $T(m)$ and the contribution of this limiting value is negligible).
Therefore, Eq.~(\ref{psi}) becomes
\begin{equation}\label{psiy}
 \psi(z)=\exp\int_{1}^{z}\frac{2z'(r-sz'^{2})}{\omega z'^{2}}dz'=z^{\alpha}{\rm e}^{\beta(1-z^{2})},
\end{equation}
where $\alpha\equiv 2r/\omega$ and $\beta\equiv s/\omega$.
The first passage time is written as
\begin{equation}\label{ty}
 T=2\int_{1/N}^{1}\frac{I(y)}{\psi(y)}dy,
\end{equation}
where it has been defined
\begin{equation} \label{iy}
 I(y)=\int_{y}^{1}\frac{\psi(z)}{Kj_{0}^{2}(z)}dz=\frac{{\rm e}^{\beta}}{\omega}\int_{y}^{1}z^{\alpha-2}{\rm e}^{-\beta z^{2}}dz,
\end{equation}
which presents a singularity when $\alpha = 1$ $(b_{0}=1\equiv b_{0,c})$. This case will be studied separately.

\subsection{Case $\alpha \neq 1$}

Integrating by parts Eq.~(\ref{iy}),
\begin{equation}
 I(y)=\frac{{\rm e}^{\beta}}{\omega}\left[\frac{{\rm e}^{-\beta}-{\rm e}^{-\beta y^{2}}y^{\alpha-1}}{\alpha-1}+2\beta\int_{y}^{1}\frac{z^{\alpha}{\rm e}^{-\beta z^{2}}}{\alpha-1}dz\right].
\end{equation}
This integral can be solved again integrating by parts, and so on, recursively,
\begin{equation}
I(y)=\frac{1}{\omega}\sum_{k=0}^{\infty}(2\beta)^{k}\frac{1-{\rm e}^{-\beta(y^{2}-1)}y^{\alpha-1+2k}}{\prod_{i=0}^{k}\alpha-1+2i}.
\end{equation}
Therefore, 
\begin{equation}
T=\frac{2}{\omega}\sum_{k=0}^{\infty}\frac{(2\beta)^{k}}{\prod_{i=0}^{k}\alpha-1+2i}\left[I_{1}(N)-I_{2}(k,N)\right], 
\end{equation}
where
\begin{eqnarray}
 I_{1}(N)&\equiv&\int_{1/N}^{1} y^{-\alpha}{\rm e}^{\beta(y^{2}-1)}dy, \\
 I_{2}(k,N)&\equiv&\int_{1/N}^{1} y^{2k-1}dy.
\end{eqnarray}
$I_{1}(N)$ is solved by parts. A recursive integration similar to the one in Eq.~(\ref{iy}) has to be performed,
\begin{equation}\label{I1}
 I_{1}(N)=\sum_{l=0}^{\infty}\frac{(-2\beta)^{l}[1-N^{\alpha-1-2l}{\rm e}^{\beta(1/N^{2}-1)}]}{\prod_{j=0}^{l}\alpha-1+2j}.
\end{equation}
On the other hand, $I_{2}(k,N)$ is easily solved
\begin{eqnarray}
I_{2}(k,N) = \left\{ \begin{array}{ll}
-\ln(N^{-1})=\ln N & \mbox{for $k=0$,} \\
\frac{1-N^{-2k}}{2k} & \mbox{for $k \geq 1$.} \\
\end{array}
\right.
\end{eqnarray}
The final expression for the first passage time is
\begin{eqnarray}
 T&=&\frac{2}{\omega}\left(\frac{I_{1}(N)-\ln N}{\alpha-1}\right)+ \nonumber \\
&+&\frac{2}{\omega}\sum_{k=1}^{\infty}\frac{(2\beta)^{k}\left[I_{1}(N)-(1-N^{-2k})/2k\right]}{\prod_{i=0}^{k}\alpha-1+2i}, \nonumber \\
\end{eqnarray}
whose asymptotic limit $N \rightarrow \infty$ has two different cases.

\subsubsection{$\alpha < 1$}

In this case, $\alpha-1-2l < 0$ when $l \geq 0$ so in $I_{1}(N)$
\begin{equation}
 1-N^{\alpha-1-2l}{\rm e}^{\beta(1/N^{2}-1)} \sim 1,
\end{equation}
which leads to
\begin{equation}
 I_{1}(N)=\sum_{l=0}^{\infty}\frac{(-2\beta)^{l}}{\prod_{j=0}^{l}1+2j-\alpha} \equiv C(\alpha,\beta).
\end{equation}
We have
\small
\begin{equation}
 T \approx \frac{2}{\omega}\left[\frac{C(\alpha,\beta)-\ln N}{\alpha-1}+\sum_{k=1}^{\infty}\frac{(2\beta)^{k}\left(C(\alpha,\beta)-(2k)^{-1}\right)}{\prod_{j=0}^{l}\alpha-1+2j}\right], 
\end{equation}
\normalsize
and finally,
\begin{equation}
 T \approx \frac{2}{\omega(\alpha-1)}\ln N.
\end{equation}

\subsubsection{$\alpha > 1$.}

Considering that $N^{\alpha-1} \gg N^{\alpha-1-2l}, \forall l>0$, only the first term is relevant in Eq.~(\ref{I1}) for 
$I_{1}(N)$. Then
\begin{equation} 
 I_{1}(N)\approx \frac{1-{\rm e}^{-\beta}N^{\alpha-1}}{1-\alpha}\approx \frac{{\rm e}^{-\beta}N^{\alpha-1}}{1-\alpha},
\end{equation}
and in the asymptotic behavior $(N \gg 1)$ of the mean escape time
\begin{eqnarray}
 T \approx K(\alpha,\beta)N^{\alpha-1}-\frac{2\ln N}{\omega(\alpha-1)} \sim N^{\alpha-1}.
\end{eqnarray}

\subsection{Case $\alpha=1$. Critical point}

We need to solve
\begin{equation}
 I(y)=\int_{y}^{1}\frac{\psi(z)}{Kj^{2}(z)}dz=\frac{{\rm e}^{\beta}}{\omega}\int_{y}^{1}y^{-1}{\rm e}^{-\beta z^{2}}dz.
\end{equation}
Expanding the exponential function and integrating, it is
\begin{equation}\label{iy2}
  I(y)=\frac{{\rm e}^{\beta}}{\omega}\left[-\ln y+\sum_{k=1}^{\infty}\frac{(-\beta)^{k}(1-2y)^{2k}}{k!2k}\right].
\end{equation}
Taking Eq.~(\ref{iy2}) into Eq.~(\ref{ty})
\begin{equation}
 T=\frac{2{\rm e}^{\beta}}{\omega}\left[I_{3}(N)+\sum_{k=1}^{\infty}\frac{(-\beta)^{k}}{k!2k}\left(I_{4}(N)+I_{5}(k,N)\right)\right],
\end{equation}
where
\begin{eqnarray}
 I_{3}(N)&=&-\int_{1/N}^{1}y^{-1}\ln y{\rm e}^{\beta(y^{2}-1)}dy, \nonumber \\
I_{4}(N)&=&\int_{1/N}^{1}y^{-1}{\rm e}^{\beta(y^{2}-1)}dy, \nonumber \\
I_{5}(k,N)&=&\int_{1/N}^{1}y^{k-1}{\rm e}^{\beta(y^{2}-1)}dy.
\end{eqnarray}
First of all, let us consider the solution of $I_{3}(N)$ integrating by parts, so that
\begin{equation}
 I_{3}(N)=\frac{(\ln N)^{2}}{2}{\rm e}^{\beta(N^{-2}-1)}+\beta\int_{1/N}^{1}(\ln y)^{2}{\rm e}^{\beta(y^{2}-1)}dy,
\end{equation}
and we obtain
\begin{equation}
 I_{3}(N)=\frac{(\ln N)^{2}}{2}{\rm e}^{\beta(N^{-2}-1)}+2\beta-O(N^{-1})+O\left(\frac{\ln N}{N}\right),
\end{equation}
which scales in the asymptotic limit as
\begin{equation}
 I_{3}(N)\sim \frac{(\ln N)^{2}}{2}{\rm e}^{-\beta}.
\end{equation}
On the other hand, the leading behavior when the size of the system is big enough $(N \gg 1)$ is
\begin{equation}
 I_{4}(N)\sim{\rm e}^{-\beta}\ln N+C_{4}(\beta).
\end{equation}
To solve the last integral,$I_{5}(k,N)$, the exponential function has to be expanded as well. It is
\begin{equation}
I_{5}(k,N)={\rm e}^{-\beta}\sum_{l=0}^{\infty}\frac{\beta^{l}}{l!(k+2l)}\left(1-N^{-2l-k}\right)\sim constant,
\end{equation}
when $N\gg 1$. It finally leads to an expression for $T$ at criticality
\small
\begin{equation}
 T \approx \frac{2{\rm e}^{-\beta}}{\omega}\left\lbrace\frac{{\rm
       e}^{-\beta}(\ln N)^{2}}{2}+
\sum_{k=1}^{\infty}\frac{(-\beta)^{k}}{k!2k}\left[{\rm e}^{-\beta}\ln N+C'_{4}(\beta)\right]\right\rbrace.
\end{equation}
\normalsize
In the limit of very large system sizes ($N \gg 1$) the mean escape time scales as
\begin{equation}
 T \sim \frac{(\ln N)^{2}}{\omega}+\frac{1}{\omega}\sum_{k=1}^{\infty}\frac{(-\beta)^{k}}{k!k}\ln N+K(\beta),
\label{quadraticpol}
\end{equation}
which asymptotically becomes
\begin{equation}
 T \sim \frac{(\ln N)^{2}}{\omega}.
\end{equation}
Summing up, the time taken by the system for reaching $m=2/N$ from an
initial condition $m=1$ is
\begin{eqnarray}
T \sim \left\{ \begin{array}{ll}
\frac{2}{\omega(\alpha-1)}\ln N & \mbox{for $\alpha < 1$,} \\
\frac{(\ln N)^{2}}{\omega} & \mbox{for $\alpha = 1$,} \\
N^{\alpha-1} & \mbox{for $\alpha > 1$.}
\end{array}
\right.
\end{eqnarray}
or in terms of the original parameters
\begin{eqnarray}
T \sim \left\{ \begin{array}{ll} 
\frac{\ln N}{b_{0}-1} & \mbox{for $b_{0} < b_{0,c}$,} \\
\frac{3(\ln N)^{2}}{\sigma^{2}} & \mbox{for $b_{0} = b_{0,c}$,} \\
N^{\frac{6(b_{0}-1)}{\sigma^{2}}} & \mbox{for $b_{0} > b_{0,c}$,}
\label{apA:final}
\end{array}
\right.
\end{eqnarray}

\section{Analytical calculations of the crossing time for the mean
  field nonlinear q-voter model}

After performing the change of variables of Eq.~(\ref{change-rho}),
the absorbing barrier is placed at $\rho=1/N$ and the reflecting one
at $\rho=1/2$, (which is the initial point). The mean first passage
time is given by \begin{equation}\label{esctime-gv}
 T(\rho=1/2)=2\int_{1/N}^{\rho=1/2}\frac{dy}{\psi(y)}\int_{y}^{1/2}\frac{\psi(z)}{Kj_{0}^{2}(z)}dz,
\end{equation}
with
\begin{equation}\label{psi-gv}
  \psi(z)=\exp\int_{1/N}^{z}dz'\frac{2A(z')+KC(z')C'(z')}{KC^{2}(z')}.
\end{equation}
We expand the polynomials in Eq.~(\ref{psi-gv}) up to second order, using Eq.~(\ref{change-function}), it is
\begin{eqnarray}
 A(\rho)+\frac{K}{2}C(\rho)C'(\rho)&\approx&\rho(r-s\rho), \nonumber \\
 KC^{2}(\rho)&\approx&\omega\rho^{2},
\end{eqnarray}
where we have defined $\omega\equiv3\sigma^{2}$,
$r=\frac{w}{2}-1+3b_{0}$ and $s=3r$.  These polynomials are similar
to the ones obtained for the Ising model, but with redefined
parameters. The integrals are done in a very similar way, and one
finally reaches the following expressions for the crossing (or ordering) time. 
\begin{eqnarray}
T \sim \left\{ \begin{array}{ll}
\frac{\ln N}{\left(3b_{0}-1\right)} & \mbox{for $b_{0} < 1/3$} \\
\frac{(\ln N)^{2}}{3\sigma^{2}} & \mbox{for $b_{0} = 1/3$} \\
N^{\frac{2(b_{0}-1/3)}{\sigma^{2}}} & \mbox{for $b_{0} > 1/3$.}
\end{array}
\right.
\end{eqnarray}

\vspace{1cm}



\begin{thebibliography}{99}
\bibitem{mf}
For a fully connected network the number of neighbours has no meaning. However, the MF limit of the model refers
to the use of the probability Eq.~(\ref{probqvoter}).
\end{thebibliography}
\end{document}